\documentclass[3p,times]{elsarticle}

\usepackage{ecrc}

\usepackage{epstopdf}

\volume{00}

\firstpage{1}

\journalname{Nuclear Physics A}

\runauth{Piotr Bo\.zek and Wojciech Broniowski}

\jid{nupha}

\jnltitlelogo{Nuclear Physics A}

\usepackage{graphicx}
\usepackage{amsmath,amssymb}
\begin{document}

\begin{frontmatter}

%% Title, authors and addresses

%% use the tnoteref command within \title for footnotes;
%% use the tnotetext command for the associated footnote;
%% use the fnref command within \author or \address for footnotes;
%% use the fntext command for the associated footnote;
%% use the corref command within \author for corresponding author footnotes;
%% use the cortext command for the associated footnote;
%% use the ead command for the email address,
%% and the form \ead[url] for the home page:
%%
%% \title{Title\tnoteref{label1}}
%% \tnotetext[label1]{}
%% \author{Name\corref{cor1}\fnref{label2}}
%% \ead{email address}
%% \ead[url]{home page}
%% \fntext[label2]{}
%% \cortext[cor1]{}
%% \address{Address\fnref{label3}}
%% \fntext[label3]{}

\title{Collective flow in small systems}

%% Single author (and collaboration) - please insert
\author[agh,ifj]{Piotr Bo\.zek}
%\ead[pbe]{piotr.bozek@ifj.edu.pl}
\author[ifj,ujk]{Wojciech Broniowski}
\address[agh]{AGH University of Science and Technology, 
Faculty of Physics and Applied Computer Science, PL-30-059 Krak\'ow, Poland}
\address[ifj]{
Institute of Nuclear Physics, PL-31342 Krak\'ow, Poland}
\address[ujk]{Institute of Physics, Jan Kochanowski University, PL-25-406 Kielce, Poland}
%\ead[wbe]{wojciech.broniowski@ifj.edu.pl}

%% For multiple authors, replace the above by:

%\author[label1]{Author1}
%\author[label2]{Author2}

%\address[label1]{Address 1}
%\address[label2]{Address 2}

\begin{abstract}
%% Text of abstract
The large density of matter in the interaction region of the proton-nucleus or
deuteron-nucleus collisions
enables the collective expansion of the fireball. Predictions of the hydrodynamic model for the asymmetric 
transverse flow are presented and compared to experimental data. \\ 
%\verb!http://www.elsevier.com/author-schemas/preparing-crc-journal-articles-with-latex!.
\end{abstract}

\begin{keyword}
%% keywords here, in the form: keyword \sep keyword
collective flow \sep hydrodynamic model \sep ultrarelativistic collisions
%% MSC codes here, in the form: \MSC code \sep code
%% or \MSC[2008] code \sep code (2000 is the default)

\end{keyword}

\end{frontmatter}

%%
%% Start line numbering here if you want
%%
% \linenumbers

%% main text

\section{Introduction}
\label{intro}
The bulk dynamics of the fireball created in ultrarelativistic heavy-ion collisions is governed by
hydrodynamics. Pressure gradients in the transverse plane lead to a rapid transverse expansion. 
The azimuthal asymmetry of  particle spectra 
%\begin{equation}
%\frac{dN}{dy p_\perp dp_\perp d\phi}  = \frac{dN}{2 \pi dy p_\perp dp_\perp }\left(1+ \sum_n 2v_n \cos\left(n(\phi-\Psi_n)\right) \right)
%\end{equation}
can be quantified with the flow coefficients $v_n$.
Inspection of the initial conditions in the high energy p-A or d-A collisions
shows that  collective 
flow should appear in such systems \cite{Bozek:2011if}. Alternatively,
saturation phenomena yield qualitatively similar angular correlations for the
emitted particles \cite{Dusling:2012cg}.
We present a selection of results obtained from the hydrodynamic model in
comparison to recent experimental 
data for p-Pb collisions at the LHC 
\cite{CMS:2012qk,Abelev:2012ola,Aad:2012gla} and d-Au collisions at RHIC
\cite{Adare:2013piz}.

\section{The fireball}

The multiplicity in p-Pb collisions at $\sqrt{s}=5.02$TeV reaches values
comparable to those in
peripheral Pb-Pb collisions at $2.76$TeV. It means that in the most violent p-Pb interactions the energy deposited in the interaction region is similar as in heavy-ion collisions. The size of the interaction region can
 be estimated in models of the initial state, e.g., the Glauber Monte Carlo
model or the IP-glasma model 
 \cite{Bozek:2011if,Bozek:2013uha,Bzdak:2013zma}. The size of the fireball in p-Pb collisions 
is smaller than in peripheral
 Pb-Pb collisions, which means that the energy density in the center of the
fireball is high and, consequently, the pressure
 gradients can generate a rapid transverse expansion. 
From the Glauber Monte Carlo model  \cite{Broniowski:2007nz} of p-Pb interactions one finds that the
number of participant nucleons 
goes beyond 20 for the most violent collisions. Additional fluctuations of the
energy deposition in each individual nucleon-nucleon interaction increase the
range of the  energies of the fireball. In fact, a simple model
which convolutes
the distribution of the participant nucleons with a negative binomial
distribution of particles emitted per
 nucleon-nucleon collision reproduces the observed distribution of the charged
tracks in p-Pb collisions
\cite{Bozek:2013uha}. Event by event fluctuations in the distribution
 of the participant nucleons in the transverse plane and in the energy
deposition yield nonzero initial eccentricities $\epsilon_n$ of the fireball
\begin{equation}
\epsilon_n e^{i n \Psi_n} =-\frac{\int r^{n+1} e^{i n\phi} \rho(r,\phi) dr d\phi}{\int r^{n+1} \rho(r,\phi) dr d\phi} \ .
\end{equation}

\begin{figure}
\begin{center}
\includegraphics*[angle=0,width=0.46 \textwidth]{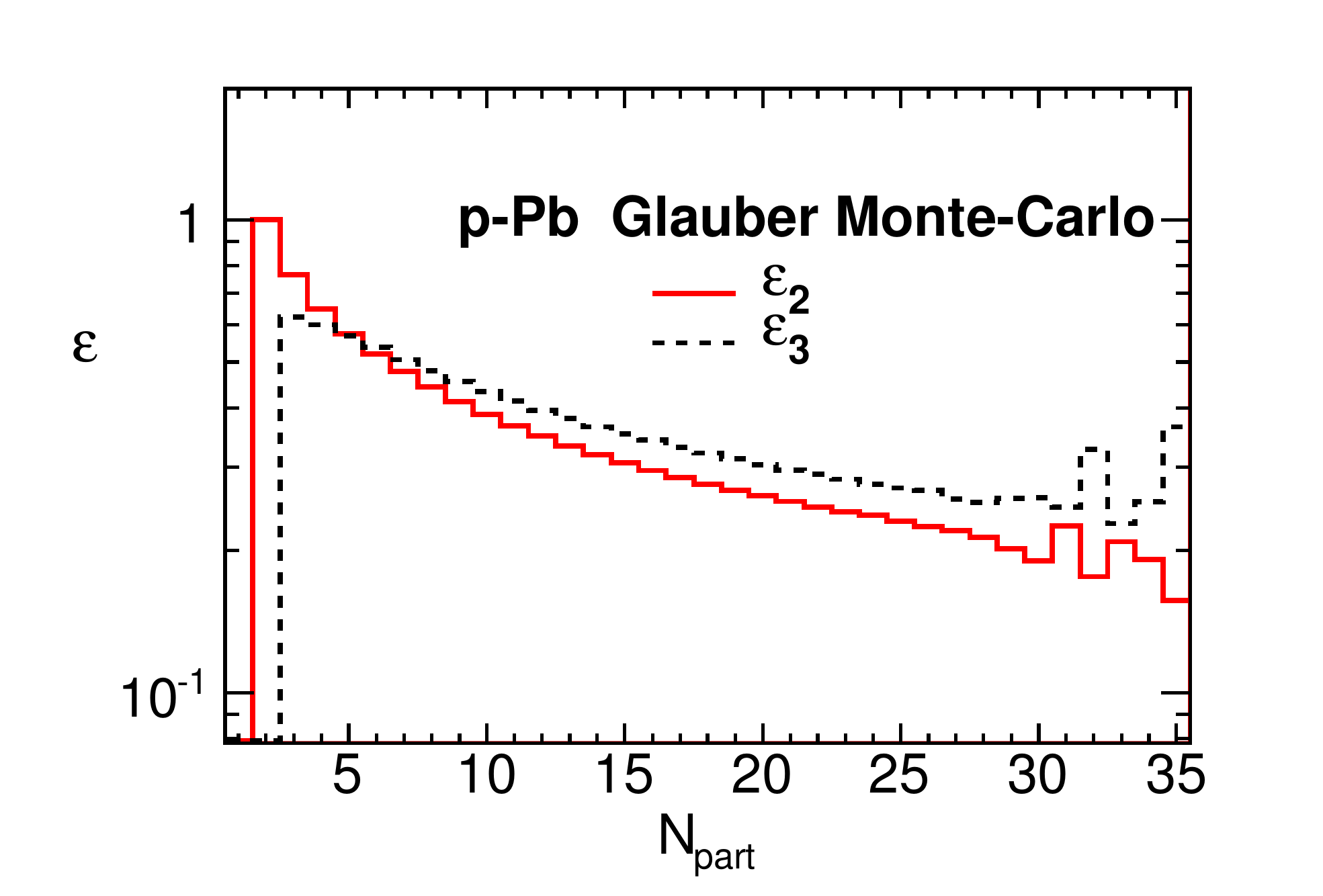}
\includegraphics*[angle=0,width=0.34 \textwidth]{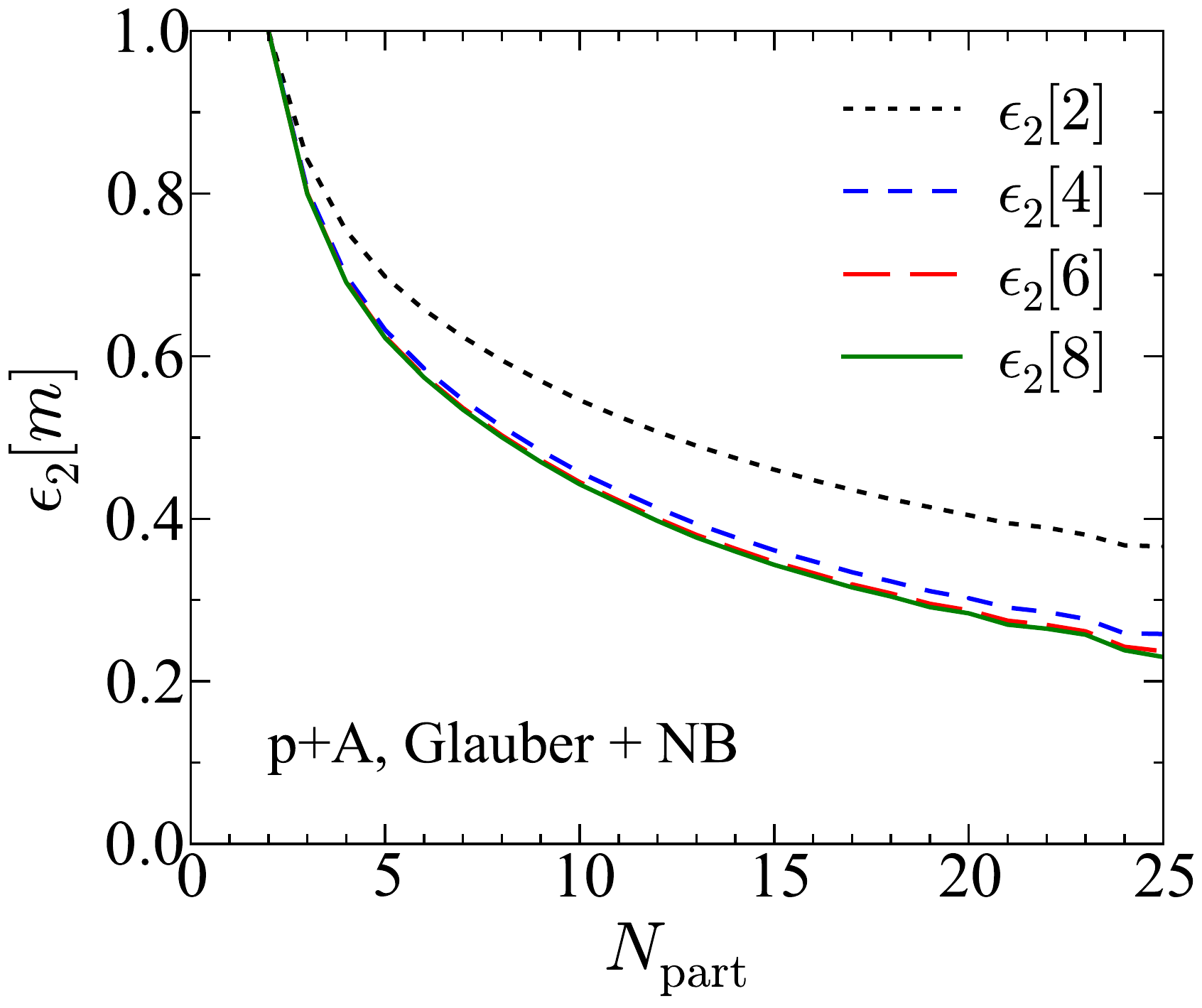} 
\end{center}
\vspace{-7mm}
\caption{(left)
Initial eccentricity and triangularity for p-Pb collisions as functions of
the number of participant nucleons 
 \cite{Bozek:2011if}. (right)~Initial eccentricity in p-Pb collisions
from higher order cumulants \cite{Bzdak:2013rya}. 
\label{fig:e2e3}} 
\end{figure}

\begin{figure}
\begin{center}
\includegraphics*[angle=0,width=0.46 \textwidth]{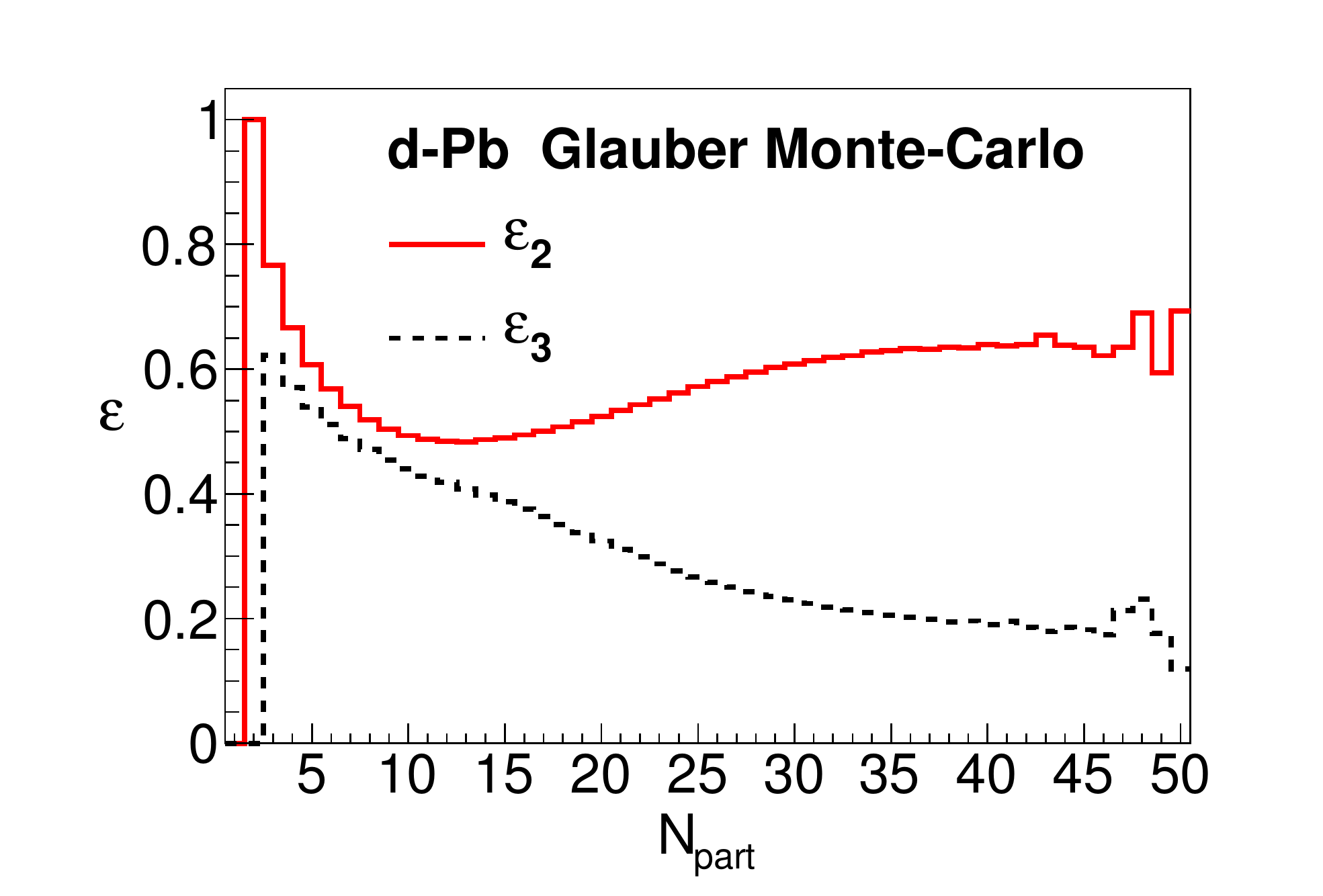}
\includegraphics*[angle=0,width=0.29 \textwidth]{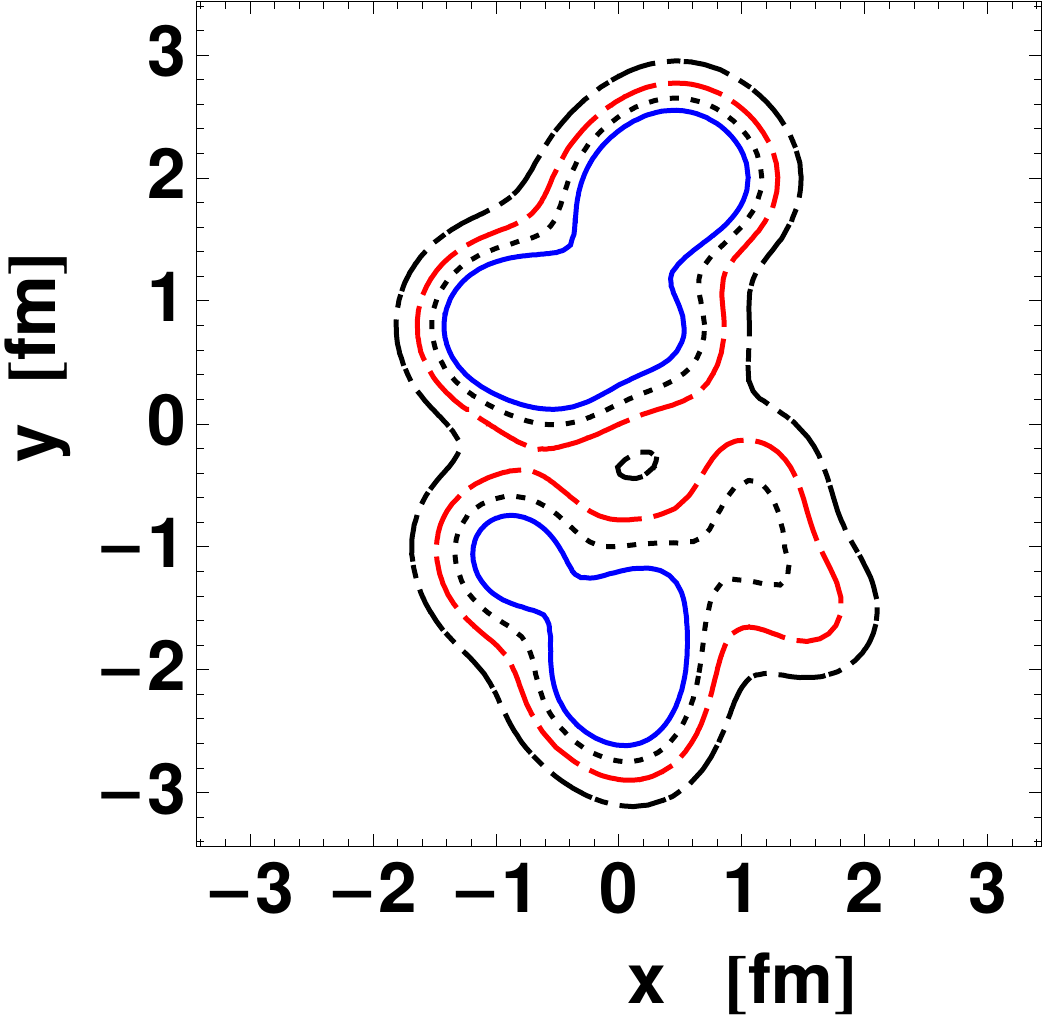} 
\end{center}
\vspace{-7mm}
\caption{(left) Initial eccentricity and triangularity for d-Pb collisions as
functions of the number of participant nucleons. 
 (right)~Initial entropy density in the transverse plane for a
typical d-Pb event \cite{Bozek:2011if}.
\label{fig:e2den}} 
\end{figure}  

In p-Pb interactions $\epsilon_2$ and $\epsilon_3$ originate solely from
the shape fluctuations 
of the fireball. These eccentricities decrease with the number of sources (left
panel of Fig.~\ref{fig:e2e3}). 
Still, the values of $\epsilon_n$ for the highest multiplicity collisions, where
the hydrodynamic  expansion takes place,
are sizable, hence the resulting $v_2$ and $v_3$ can be measured. The
contributions from fluctuations  
and from the average asymmetry to the eccentricity can be separated using multiparticle cumulant measures for the 
flow coefficients \cite{Borghini:2000sa}. In the limit of a large number of
sources the higher order 
cumulants should vanish, as they measure the shape asymmetry. 
Since in p-Pb collisions the number of participant
nucleons is limited, 
the Glauber Monte Carlo model gives a nonzero value of the higher order
cumulants $\epsilon_n$
(right panel of Fig. \ref{fig:e2e3}) 
\cite{Bzdak:2013rya,Yan:2013laa}, with the elliptic and triangular
flow coefficients 
following approximately the relation $v_n\{2\}>v_n\{4\}\simeq v_n\{6\} \simeq
v_n\{8\}$,
 as observed experimentally 
\cite{cmswiki2}.

When a deuteron hits  a large nucleus side-wise, the azimuthal asymmetry is
large 
due to the  deformation of the intrinsic wave function of the deuteron
\cite{Bozek:2011if}. The shape of the
 fireball in the transverse plane reflects the large  eccentricity of the density profile 
(right panel of Fig. \ref{fig:e2den}). By triggering on events with a large number of 
participant nucleons,  configurations with a large $\epsilon_2$  are selected
 (left panel of Fig. \ref{fig:e2den}). The same argument applies to tritium-Au or $^3$He-Au
collisions  \cite{Sickles:2013mua,Nagle:2013lja}. High multiplicity events
correspond to configurations where the smaller
nucleus hits the larger nucleus  with a large separation between the three
nucleons, resulting in large triangularity.
The $^{12}$C  nucleus exhibits strong nuclear correlations, as it consists of
three $\alpha$-like clusters. The fireball formed in high multiplicity
collisions of such nuclei with a large target nucleus exhibits significant
geometric triangularity that should be 
observable as an increase of the triangular flow for the highest multiplicity
events \cite{Broniowski:2013dia}.

\section{Results of the hydrodynamic model}

\begin{figure}
\begin{center}
\includegraphics*[angle=0,width=0.45 \textwidth]{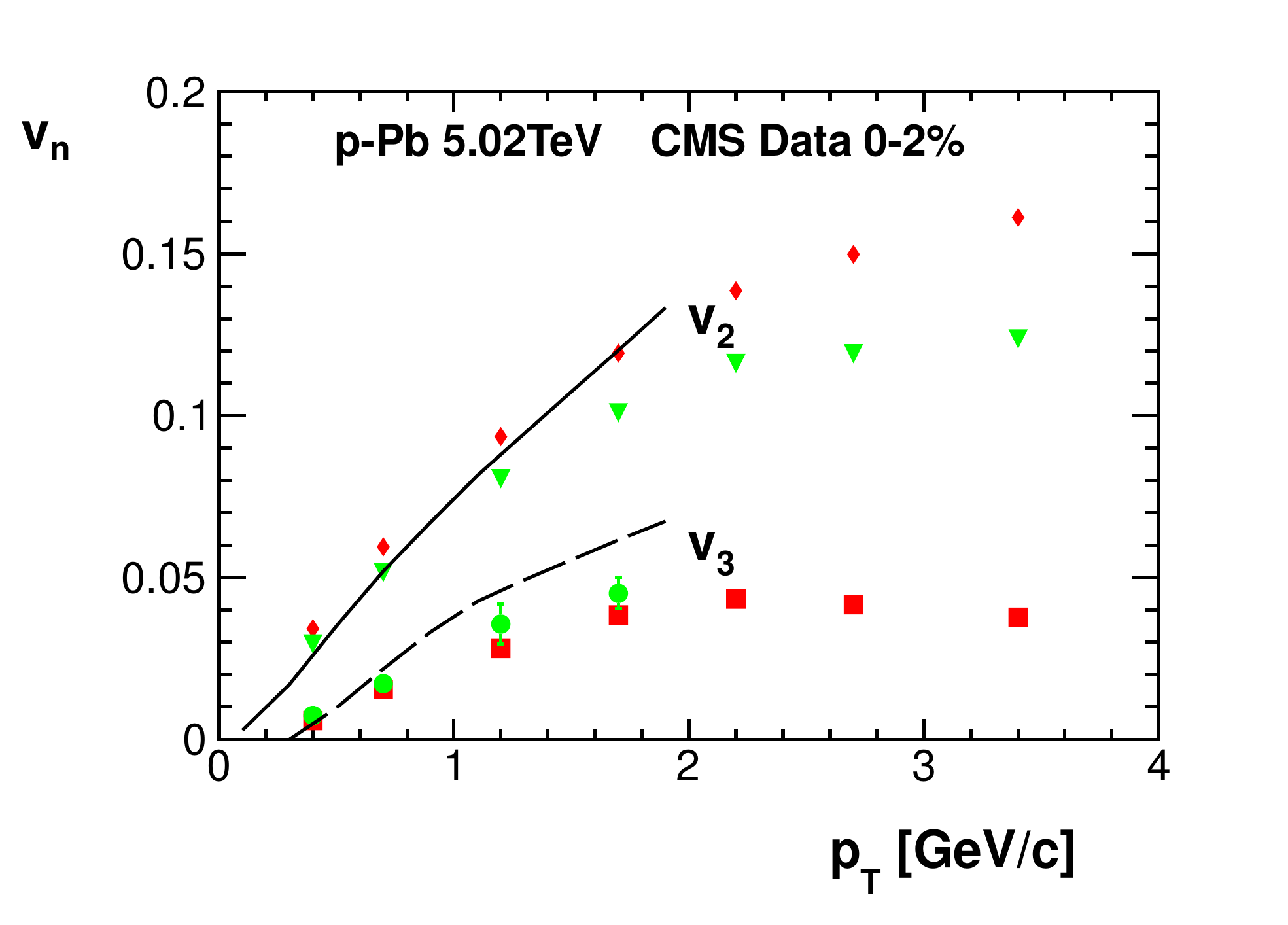}
\includegraphics*[angle=0,width=0.45 \textwidth]{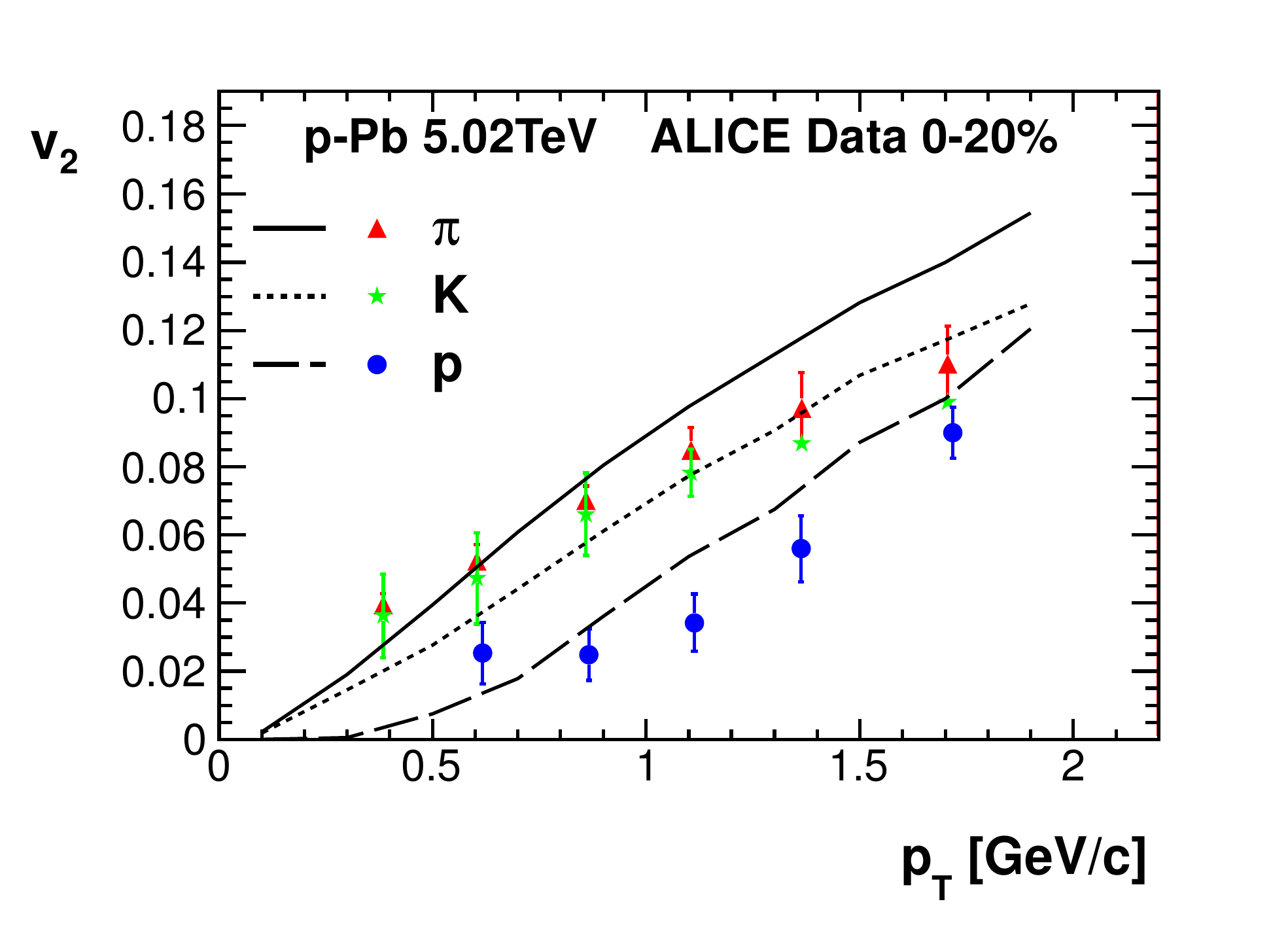} 
\end{center}
\vspace{-7mm}
\caption{(left)
Elliptic and triangular flow coefficients of charged particles in p-Pb collisions from the hydrodynamic model 
compared 
to the CMS data \cite{Chatrchyan:2013nka}.
 (right) Elliptic flow coefficient of identified particles from the
hydrodynamic model compared to the ALICE 
data \cite{ABELEV:2013wsa}.
\label{fig:v2v3}} 
\end{figure}

The expansion of the fireball 
is described using  3+1-dimensional viscous hydrodynamic simulations with shear and bulk viscosity
\cite{Bozek:2009dw}. 
At the freeze-out temperature of $150$MeV, hadrons are emitted statistically and resonance decays are accounted for \cite{Chojnacki:2011hb}. The main source of uncertainty in the model calculations resides in the assumptions on 
the initial conditions. The details of the hydrodynamic stage involve additional parameter dependence. Nevertheless, most of the calculations predict flow coefficients consistent with experiment \cite{Bozek:2013uha,Qin:2013bha,Nagle:2013lja,Werner:2013ipa,Kozlov:2014fqa,Bzdak:2014dia}, with the exception of the calculation using IP-glasma initial conditions \cite{Schenke:2014zha}.

\begin{figure}[t]
\begin{center}
\includegraphics*[angle=0,width=0.500 \textwidth]{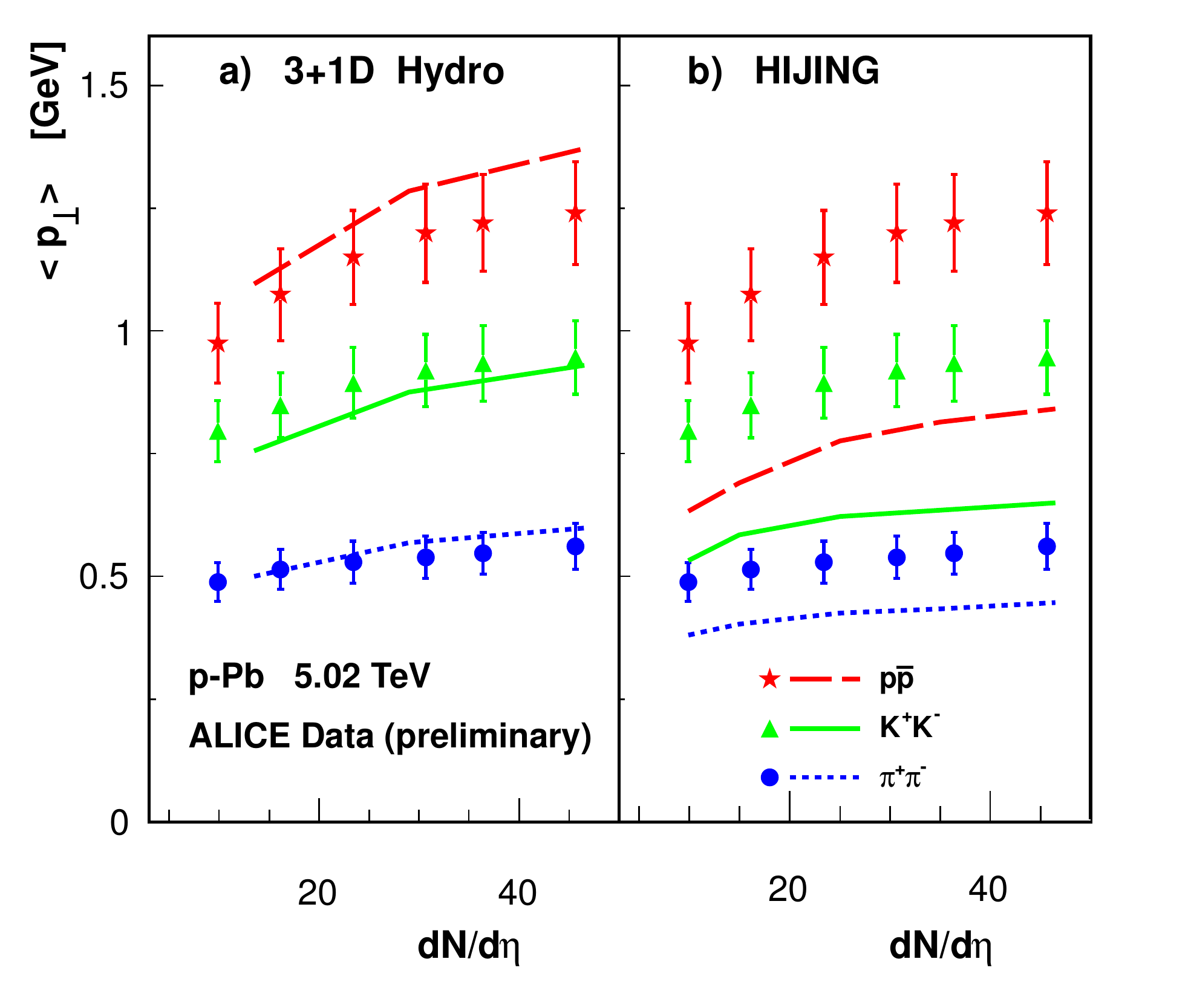}
\end{center}
\vspace{-7mm}
\caption{Average transverse momentum of identified particles  in p-Pb
collisions, data from 
ALICE Collaboration \cite{Abelev:2013haa}, compared to the results of the 
HIJING model and 
 of the viscous hydrodynamics. 
\label{fig:pt}} 
\end{figure}

The resulting elliptic and triangular flow coefficients for charged particles
are shown in Fig. \ref{fig:v2v3}. The calculation
 describes reasonably well the measured flow asymmetry. The observed elliptic
and triangular flow in p-Pb collisions 
is similar as in peripheral Pb-Pb collisions, suggesting a common
origin. Hydrodynamic calculations in the two systems 
explain uniformly these observations via formation of the collective flow. We
note that the presence of nonzero harmonic flow coefficients  
generates ridge-like structures in the two particle correlation function in
the relative
 azimuthal angle and pseudorapidity 
\cite{Bozek:2012gr} .
 The elliptic flow in p-Pb collisions for identified particles 
shows a splitting
 between particles of different mass (Fig. \ref{fig:v2v3} right panel). 
This behavior arises naturally in the framework of the hydrodynamic 
model \cite{Bozek:2013ska}.
A strong elliptic flow is observed in d-Au collisions at RHIC energies \cite{Adare:2013piz}, consistent with 
earlier predictions of the hydrodynamic model. 
The presence of the collective transverse flow increases 
the average transverse momentum of the emitted particles. The transverse
momentum of
particles emitted in p-Pb collisions is larger than in models consisting of a convolution of nucleon-nucleon collisions 
(Fig. \ref{fig:pt} right panel) \cite{Bzdak:2013lva}. The hydrodynamic model predicts a larger transverse flow for heavier
particles ( Fig. \ref{fig:pt} left  panel) \cite{Bozek:2013ska} and thus
explains the mass hierarchy observed experimentally.
For the asymmetric p-Pb or d-Au collisions, the hydrodynamic model predicts stronger collectivity, i.e.
 larger $v_n$ and larger transverse momenta, on the nucleus-going side.

\section{Conclusions}

The high energy density in the fireball created in high-multiplicity p-Pb or
d-Au collisions at ultrarelativistic energies implies 
the possibility of a collective expansion of such a small system. Hydrodynamic calculations are in semi-quantitative 
agreement with experimental measurements of  the harmonic flow coefficients and the average transverse momenta of emitted particles.

\bigskip

\noindent
{\bf Acknowledgments:} 

Supported by National Science Centre grant 
DEC-2012/05/B/ST2/02528 and by PL-Grid Infrastructure.

\bibliography{../../../hydr}
%\bibliography{../hydr}

%% The Appendices part is started with the command \appendix;
%% appendix sections are then done as normal sections
%% \appendix

%% \section{}
%% \label{}

%% References
%%
%% Following citation commands can be used in the body text:
%% Usage of \cite is as follows:
%%   \cite{key}         ==>>  [#]
%%   \cite[chap. 2]{key} ==>> [#, chap. 2]
%%

%% References with BibTeX database:

%\bibliographystyle{elsarticle-num}
%\bibliography{<your-bib-database>}

%% Authors are advised to use a BibTeX database file for their reference list.
%% The provided style file elsarticle-num.bst formats references in the required Procedia style

%% For references without a BibTeX database:

\end{document}